\documentclass[twocolumn,showpacs,aps,prb]{revtex4-1}

\usepackage{amssymb,amsmath}
\usepackage[dvips]{graphicx}

\begin{document}

\title{Composition-tuned smeared phase transitions}

\author{Fawaz Hrahsheh}
\author{David Nozadze}
\author{Thomas Vojta}
\affiliation{Department of Physics, Missouri University of Science and Technology, Rolla, MO 65409, USA}

\begin{abstract}
Phase transitions in random systems are smeared if individual spatial regions
can order independently of the bulk system. In this paper, we study such smeared
phase transitions (both classical and quantum) in substitutional alloys A$_{1-x}$B$_x$ that can be tuned from
an ordered phase at composition $x=0$ to a disordered phase at $x=1$. We show that
the ordered phase develops a pronounced tail that extends over all compositions
$x<1$. Using optimal fluctuation theory, we derive the composition dependence of
the order parameter and other quantities in the tail of the smeared phase transition.
We also compare our results to computer simulations of a toy model, and we discuss
experiments.
\end{abstract}

\date{\today}

\maketitle

\section{Introduction}

When a phase transition occurs in a randomly disordered system,
one of the most basic questions to ask is whether the transition
is still sharp, i.e., associated with a singularity in the free energy.
Naively, one might expect that random disorder rounds or smears any critical point
because different spatial regions undergo the transition at different
values of the control parameter. This expectation turns out to be mistaken,
as classical (thermal) continuous phase transitions generically remain sharp in the
presence of weak randomness. The reason is that a finite-size region cannot
undergo a true phase transition at any nonzero temperature because its partition
function must be analytic. Thus, true static long-range order can only be
established via a collective phenomenon in the entire system
(see, e.g., Ref.\ \onlinecite{Grinstein85} for a pedagogical discussion).

Recent work has established, however, that some phase transitions are indeed
smeared by random disorder. This can happen at zero-temperature quantum phase transitions
when the order parameter fluctuations are overdamped because they are coupled
to an (infinite) heat bath.\cite{Vojta03a,HoyosVojta08} As the damping hampers the
dynamics, sufficiently large
but finite-size regions can undergo the phase transition independently from the
bulk system. Once several such regions have developed static order, their local order
parameters can be aligned by an \emph{infinitesimally small} mutual interaction.
Thus, global order develops gradually, and the global phase transition is smeared.
Classical thermal phase transitions can also be smeared provided the disorder is perfectly
correlated in at least two dimensions. In these cases, individual ``slabs'' of finite
thickness undergo the phase transition independently of the bulk
system.\cite{Vojta03b,SknepnekVojta04}

The existing theoretical work on smeared phase transitions focuses on situations
in which a sample with some fixed degree of randomness is tuned through the transition
by changing the temperature (for classical transitions) or the appropriate quantum
control parameter such as pressure or magnetic field (for quantum phase transitions).
However, many experiments are performed on substitutional alloys such as CePd$_{1-x}$Rh$_x$ or
Sr$_{1-x}$Ca$_x$RuO$_3$. These materials can be tuned from an ordered phase (ferromagnetic for the two
examples) at composition $x=0$ to a disordered phase at $x=1$ while keeping the temperature
and other external parameters fixed, i.e., they undergo a phase transition as a function
of composition. The composition parameter $x$ actually plays a dual role in these
transitions. On the one hand, $x$ is the control parameter of the phase transition.
On the other hand, changing $x$ also changes the degree of randomness. If such a composition-tuned phase
transition is smeared, its behavior can therefore be expected to be different than that
of smeared transitions occurring at fixed randomness.

In this paper, we investigate the properties of composition-tuned smeared phase transitions
in substitutional alloys of the type A$_{1-x}$B$_x$. We show that the ordered phase extends
over the entire composition range $x<1$, and we derive the behavior of the system in the
tail of the smeared transition. Our paper is organized as follows. In Sec.\ \ref{sec:OFT}, we consider a
smeared quantum phase transition in an itinerant magnet. We use optimal fluctuation
theory to derive the composition dependence of the order parameter, the phase boundary, and other
quantities. In Section \ref{sec:classical} we briefly discuss how the theory is modified
for smeared classical transitions in systems with correlated disorder.
Section \ref{sec:simulations} is devoted to computer simulations of a toy model
that illustrate and confirm our theory. We conclude in Sec.\ \ref{sec:conclusions} by comparing
composition-tuned smeared transitions with those occurring at fixed randomness.
We also discuss experiments.

\section{Smeared quantum phase transition}
\label{sec:OFT}
\subsection{Model and phase diagram}
\label{subsec:model}

In this section we investigate the ferromagnetic or antiferromagnetic quantum phase transition
of itinerant electrons with Ising order parameter symmetry. In the absence of quenched
randomness, the Landau-Ginzburg-Wilson free energy functional of this transition
in $d$ space dimensions reads \cite{Hertz76,Millis93}
\begin{equation}
S=\int dy dz ~\psi(y)\Gamma(y,z)\psi(z)+ u \int dy~\psi^{4}(y)\,.
\label{eq:clean-action}
\end{equation}
Here, $\psi$ is a scalar order parameter field, $y\equiv(\mathbf{y},\tau)$ comprises imaginary
time $\tau$ and $d$-dimensional spatial position
$\mathbf{y}$, $\int dy\equiv\int d\mathbf{y}\int_{0}^{1/T}{\rm d}\tau$, and $u$
is the standard quartic coefficient. $\Gamma(y,z)$ denotes the bare inverse propagator
(two-point vertex) whose Fourier transform reads
\begin{equation}
\Gamma(\mathbf{q},\omega_{n})=r+\xi_{0}^{2}\mathbf{q}^{2}+\gamma_0(\mathbf{q})\left|\omega_{n}\right|~.
\label{eq:bare_Gamma}
\end{equation}
Here, $r$ is the distance from criticality,\footnote{Strictly, one needs to distinguish the bare distance
from criticality that appears in (\ref{eq:bare_Gamma}) from the renormalized one that measures the distance
from the true critical point. We suppress this difference because it is unimportant for our purposes.}
 $\xi_{0}$ is a
microscopic length scale, and $\omega_{n}$ is a Matsubara frequency. The dynamical part of
$\Gamma(\mathbf{q},\omega_{n})$ is proportional to $|\omega_{n}|$. This overdamped dynamics
reflects the Ohmic dissipation caused
by the coupling between the order parameter fluctuations and the gapless fermionic excitations in an itinerant system.
The damping coefficient $\gamma_0(\mathbf{q})$ is $\mathbf{q}$-independent for an antiferromagnetic
transition but proportional to $1/|\mathbf{q}|$ or $1/|\mathbf{q}|^2$ for ballistic and diffusive ferromagnets,
respectively.

We now consider two materials A and B. Substance A is in the magnetic phase,
implying a negative distance from criticality, $r_A<0$, while substance B is nonmagnetic
with $r_B>0$. By randomly substituting B-atoms for the A-atoms to form a binary alloy
A$_{1-x}$B$_x$, we can drive the system through a composition-driven magnetic quantum phase
transition.

A crucial role in this transition is played by rare A-rich spatial regions. They can
be locally in the magnetic phase even if the bulk system is nonmagnetic.
In the presence of Ohmic dissipation, the low-energy physics of each
such region is equivalent to that of a dissipative two-level system which is known
to undergo, with increasing dissipation strength, a phase transition from a fluctuating
to a localized phase.\cite{LCDFGZ87}
Therefore, the quantum dynamics of sufficiently large rare regions completely
freezes,\cite{MillisMorrSchmalian01} and they behave as classical superspins.
At zero temperature, these classical superspins can be aligned by an \emph{infinitesimally small}
residual interaction which is always present as they are coupled via the
fluctuations of the paramagnetic bulk system.
The order parameter is thus spatially very inhomogeneous, but its average
is nonzero for any $x<1$ implying
that the  global quantum phase transition is smeared by the disorder inherent
in the random positions of the A and B atoms.\cite{Vojta03a,Vojta06,Vojta10}

\begin{figure}[t]
\includegraphics[width=7.5cm,clip]{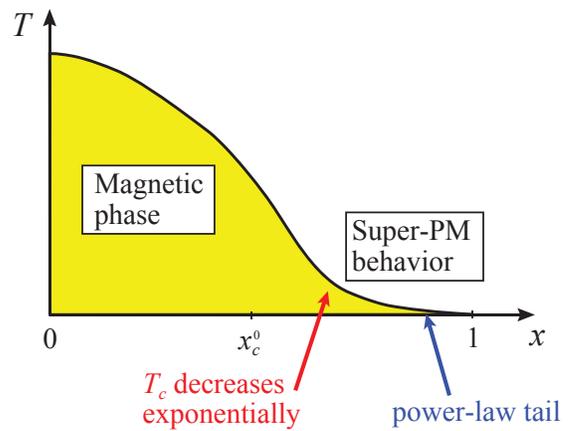}
\caption{(Color online) Schematic temperature-composition phase diagram of a binary alloy A$_{1-x}$B$_x$
    displaying a smeared quantum phase transition. In the tail of the magnetic phase,
    which stretches all the way to $x=1$, the rare regions are aligned. Above $T_c$,
    they act as independent classical moments, resulting in super-paramagnetic (PM)
    behavior. $x_c^0$ marks the critical composition in average potential approximation
    defined in (\ref{eq:xc0}).}
\label{fig:pd}
\end{figure}

At small but nonzero temperatures, the static magnetic order on the rare regions
is destroyed, and a finite interaction of the order of the temperature is necessary
to align them. This restores a sharp phase transition at some transition temperature
$T_c(x)$ which rapidly decreases with increasing $x$ but reaches zero only at $x=1$.
If the temperature is raised above $T_c$, the locally ordered rare regions act
as independent classical moments, leading to super-paramagnetic behavior.
A sketch of the resulting phase diagram is shown in Fig.\ \ref{fig:pd}.

\subsection{Optimal fluctuation theory}
\label{subsec:OFT}

In this section, we use optimal fluctuation theory \cite{Lifshitz64,HalperinLax66}
to derive the properties of the 'tail' of the smeared quantum phase transition.
This is the composition range where a few rare regions have developed static magnetic
order but their density is so small that they are very weakly coupled.

A crude estimate of the transition point in the binary alloy A$_{1-x}$B$_x$ can
be obtained by simply averaging the distance from criticality,
$r_{\rm av} = (1-x) r_A + x r_B$. The transition point corresponds to $r_{\rm av} = 0$.
This gives the critical composition in ``average potential approximation,''
\begin{equation}
x_c^0 = -r_A/(r_B - r_A)~.
\label{eq:xc0}
\end{equation}

Let us now consider a single A-rich rare region of linear size  $L_{RR}$
embedded in a nonmagnetic bulk sample. If the concentration $x_{\rm loc}$ of B atoms
in this region is below some critical concentration $x_c(L_{RR})$, the region will develop local
magnetic order. The value of the critical concentration follows straightforwardly
from finite-size scaling,\cite{Barber_review83,Cardy_book88}
\begin{equation}
x_c(L_{RR}) = x_c^0 - D L_{RR}^{-\phi}~,
\label{eq:fss}
\end{equation}
where $\phi$ is the finite-size shift exponent and $D$ is a constant. Within mean-field theory
(which should be qualitatively correct in our case because the clean transition is above
its upper critical dimension\cite{Hertz76}), one finds $\phi=2$ and $D=\xi_0^2/(r_B -r_A)$.
Since $x_c(L_{RR})$ must be positive, (\ref{eq:fss}) implies that a rare region
needs to be larger than $L_{\rm min} = (D/x_c^0)^{1/\phi}$ to develop local magnetic order.

As the last ingredient of our optimal fluctuation theory, we now analyze the random distribution
of the atoms in the sample. For simplicity, we assume that the lattice sites are occupied
\emph{independently} by either A or B atoms with probabilities $1-x$ and $x$, respectively.
Modifications due to deviations from a pure random distribution (i.e., clustering) will be discussed
in the concluding section \ref{sec:conclusions}.
The probability of finding $N_B =N x_{\rm loc}$ sites occupied by B atoms in a spatial region
with a total of $N \sim L_{RR}^d$ sites
is given by the binomial distribution
\begin{equation}
P(N,x_{\rm loc}) = \binom{N}{N_B} (1-x)^{N-N_B} x^{N_B}~.
\label{eq:binomial}
\end{equation}
We are interested in the regime $x>x_c^0$ where the bulk system will not be magnetically ordered
but $x_{\rm loc} = N_B/N < x_c (L_{RR})$ such that local order is possible in the region considered.

To estimate the total zero-temperature order parameter $M$ in the tail of the smeared transition (where the rare regions are
very weakly coupled), we can simply sum over all rare regions displaying local order
\begin{equation}
M \sim \int_{L_{\rm min}}^\infty dL_{RR} \int_0^{x_c(L_{RR})} dx_{\rm loc} \, m(N,x_{\rm loc}) P(N,x_{\rm loc})~.
\label{eq:M-integral}
\end{equation}
Here, $m(N,x_{\rm loc})$ is the order parameter of a single region of $N$ sites and local composition $x_{\rm loc}$;
and we have suppressed a combinatorial prefactor. We now analyze this integral in two parameter regions,
(i) the regime where $x$ is somewhat larger than $x_c^0$ but not by too much, and (ii) the far tail of
the transition at $x\to 1$.

If $x$ is not much larger than $x_c^0$, the rare regions are expected to be large, and we can approximate the
binomial distribution (\ref{eq:binomial}) by a Gaussian,
\begin{equation}
P(N,x_{\rm loc}) = \frac 1 {\sqrt{2\pi N (1-N)} }\exp\left [ - N \frac{(x_{\rm loc}-x)^2}{2x(1-x)} \right ]
\label{eq:Gaussian}
\end{equation}
To exponential accuracy in $x$, the integral (\ref{eq:M-integral}) can now be easily performed
in saddle point approximation. Neglecting $m(N,x_{\rm loc})$, which only modifies power-law prefactors,
we find that large rare regions of size $L_{RR}^\phi = D(2\phi-d)/[d(x-x_c^0)]$ and maximum possible
B-concentration $x_{\rm loc} = x_c^0-D L_{RR}^{-\phi}$ dominate the integral. Inserting these saddle point
values into the integrand yields the composition dependence of the order parameter
as\footnote{This result is valid for
$d<2\phi$ which is fulfilled for our transition. In the opposite case, the integral over $L_{RR}$ is dominated
by its lower bound, resulting in a purely Gaussian dependence of $M$ on $x-x_c^0$.}
\begin{equation}
M \sim \exp\left [ -C \frac {(x-x_c^0)^{2-d/\phi}}{x(1-x)}\right ]
\label{eq:M-exponential}
\end{equation}
where $C=2(D/d)^{d/\phi}(2\phi-d)^{d/\phi-2}\phi^2$ is a non-universal constant.

Let us now analyze the far tail of the smeared transition, $x\to 1$. In this regime, the binomial distribution
cannot be approximated by a Gaussian. Nonetheless, the integral (\ref{eq:M-integral}) can be estimated in saddle-point
approximation. We find that for $x\to 1$, the integral is dominated by pure-A regions of the minimum size that
permits local magnetic order. This means $L_{RR} = L_{\rm min} = (D/x_c^0)^{1/\phi}$ and $x_{\rm loc}=0$.
Inserting these values into the integrand of (\ref{eq:M-integral}), we find that the leading composition dependence
of the order parameter in the limit $x\to 1$ is given by a non-universal power law,
\begin{equation}
M \sim (1-x)^{L_{\rm min}^d} = (1-x)^{(D/x_c^0)^{d/\phi}}~.
\label{eq:M-power}
\end{equation}
We thus find that $M$  is nonzero in the entire composition range $0\le x<1$, illustrating the
notion of a smeared quantum phase transition.

So far, we have focused on the zero-temperature order parameter. Other quantities can be found in an
analogous manner. Let us, for example, determine the phase boundary, i.e., the composition dependence of the critical
temperature $T_c$. As was discussed in Sec.\ \ref{subsec:model}, the static magnetism of the rare regions
is destroyed at nonzero temperatures. Consequently, magnetic long-range order in the sample can only develop, if the rare
regions are coupled by an interaction of the order of the temperature.
The typical distance between neighboring locally ordered rare regions can be estimated from their
density, $\rho$, as $r_{\rm typ} \sim \rho^{-1/d} \sim M^{-1/d}$.
Within the
Landau-Ginzburg-Wilson theory (\ref{eq:clean-action},\ref{eq:bare_Gamma}), the interaction between
two rare regions drops off exponentially with their distance $r$,
$E_{\rm int} \sim \exp(-r/\xi_b)$, where $\xi_b$ is the bulk correlation length.  This leads to a
double-exponential dependence of $T_c$ on $x$ for compositions somewhat above $x_c^0$, i.e.,
$\ln(1/T_c) \sim \exp\{C(x-x_c^0)^{2-d/\phi}/[dx(1-x)]\}$. For $x\to 1$, we find
$\ln(1/T_c) \sim (1-x)^{-L_{\rm min}^d/d}$. However, in a real metallic magnet, the locally ordered
rare regions are coupled by an RKKY-type interaction that decays as a power law with distance,
$E_{\rm int} \sim r^{-d}$, rather than exponentially.\cite{DobrosavljevicMiranda05}
(This interaction is not contained in the long-wavelength expansion implied in (\ref{eq:bare_Gamma}).)
Therefore, the composition dependence of the critical temperature takes the same form
as that of the magnetization,
\begin{equation}
T_c \sim \exp\left [ -C \frac {(x-x_c^0)^{2-d/\phi}}{x(1-x)}\right ]
\label{eq:Tc-exponential}
\end{equation}
for compositions somewhat above $x_c^0$ and
\begin{equation}
T_c \sim (1-x)^{L_{\rm min}^d} = (1-x)^{(D/x_c^0)^{d/\phi}}
\label{eq:Tc-power}
\end{equation}
in the far tail of the smeared transition, $x\to 1$.

We now turn to the order parameter susceptibility. It consists of two different
contributions, one from the paramagnetic bulk system and one from the locally ordered rare regions.
The bulk system provides a finite, non-critical background throughout the tail of the smeared
transition. Let us discuss the rare region contribution in more detail. At zero temperature,
the total order parameter $M$ is nonzero for all $x<1$. The rare regions therefore always
feel a symmetry-breaking effective field which cuts off any possible divergence of
their susceptibilities. We conclude that the zero-temperature susceptibility does not
diverge anywhere in the tail of the smeared transition. If the temperature is raised
above $T_c$, the relative alignment of the rare regions is lost, and they behave as
independent large (classical) moments, leading to a super-paramagnetic temperature
dependence of the susceptibility, $\chi \sim 1/T$ (see Fig.\ \ref{fig:pd}). At even higher temperatures, when the damping of the
quantum dynamics becomes unimportant, we expect the usual non-universal quantum Griffiths
power-laws, $\chi \sim T^{\lambda-1}$, where $\lambda$ is the Griffiths
exponent.\cite{CastroNetoJones00,Vojta06,Vojta10}

\section{Smeared classical phase transition}
\label{sec:classical}

Classical (thermal) phase transitions with uncorrelated disorder cannot be smeared because
all rare regions are of finite size and can thus not undergo a true phase transition
at any nonzero temperature. However, perfect disorder correlations in one or more dimensions
lead to rare regions that are infinitely extended in the thermodynamic limit. If the number of
correlated dimensions is high enough, these infinitely large rare regions can undergo
the phase transition independently of the bulk system, leading to a smearing of the global
phase transition.\cite{Vojta03b} This happens, for example, in a randomly layered Ising magnet,
i.e., an Ising model with disorder correlated in two dimensions.\cite{SknepnekVojta04}

In this section, we discuss how the theory of Sec.\ \ref{sec:OFT} is modified for these
smeared classical phase transitions. For definiteness, we consider a classical Landau-Ginzburg-Wilson
free energy in $d$ dimensions,
\begin{equation}
S=\int d\mathbf{y} ~\psi(\mathbf{y})[r-\partial^2_{\mathbf{y}}]\psi(\mathbf{y})+ u \int d\mathbf{y}~\psi^{4}(\mathbf{y})\,.
\label{eq:classical-action}
\end{equation}
As in the quantum case, we now consider a binary ``alloy'' A$_{1-x}$B$_x$ of two materials A and B.
The atoms are arranged randomly in $d_\perp$ dimensions, while they are perfectly correlated
in $d_\parallel =d-d_\perp$ dimensions. For example, if $d_\perp=1$ and $d_\parallel=2$,
the system would consist of a random sequence of layers, each made up of only A atoms or only B atoms.

If the correlated dimension $d_\parallel$ is sufficiently large, the ``alloy'' undergoes
a smeared classical phase transition as the composition $x$ is tuned from 0 to 1 at
a (fixed) temperature at which material A is magnetically ordered, $r_A <0 $, while material B is
in the nonmagnetic phase, $r_B > 0$.
The optimal fluctuation theory for the behavior in the tail of the smeared transition can
be developed along the same lines as the theory in Sec.\ \ref{sec:OFT}. The only important difference
stems from the fact that the randomness is restricted to $d_\perp$ dimensions. The dimensionality
$d$ in eqs.\ (\ref{eq:M-exponential}) and (\ref{eq:M-power}) therefore needs to
be replaced by $d_\perp$, leading to
\begin{equation}
M \sim \exp\left [ -C \frac {(x-x_c^0)^{2-d_\perp/\phi}}{x(1-x)}\right ]
\label{eq:M-exponential-classical}
\end{equation}
for compositions somewhat above $x_c^0$ and
\begin{equation}
M \sim (1-x)^{L_{\rm min}^{d_\perp}} = (1-x)^{(D/x_c^0)^{d_{\perp}/\phi}}~
\label{eq:M-power-classical}
\end{equation}
for $x\to 1$. The same substitution of $d$ by $d_\perp$ was also found for smeared classical
transitions tuned by temperature rather than composition.\cite{Vojta03b}

\section{Computer simulations}
\label{sec:simulations}

To verify the predictions of the optimal fluctuation theory in Sec.\ \ref{sec:OFT}
and to illustrate our results, we have performed  computer simulations of a toy
model, viz., a classical Ising model with $d$ space-like dimensions and one time-like
dimension. The interactions are between nearest neighbors in the space-like
directions but infinite-ranged in the time-like ones. This $(d+1)$-dimensional toy model
retains the possibility of static order on the rare regions (which is
crucial for the transition being smeared) but permits system sizes large
enough to study exponentially rare events. The Hamiltonian reads
\begin{equation}
H= - \frac 1 L_\tau\sum_{\langle{\mathbf{y},\mathbf{z}}\rangle,\tau,\tau'}   S_{\mathbf{y},\tau}
S_{\mathbf{z},\tau'}
   - \frac 1 L_\tau\sum_{\mathbf{y},\tau,\tau'} J_{\mathbf{y}} S_{\mathbf{y},\tau} S_{\mathbf{y},\tau'}
\label{eq:toy}
\end{equation}
Here $\mathbf{y}$ and $\mathbf{z}$ are $d$-dimensional space-like coordinates
and $\tau$ is the time-like coordinate.
$L_\tau$ is the system size in time direction and $\langle {\mathbf{y},\mathbf{z}} \rangle$
denotes pairs of nearest neighbors on the hyper-cubic lattice in space. $J_{\mathbf{y}}$ is a quenched random
variable having the binary distribution $P(J) = (1-x)~ \delta(J-J_h) + x~ \delta(J-J_l)$
with $J_h>J_l$. In
this classical model $L_\tau$ plays the role of the inverse temperature in the
corresponding quantum system and the classical temperature plays the role of
the quantum tuning parameter. Because the interaction is infinite-ranged in time,
the time-like dimension can be treated in mean-field theory. For
$L_\tau\to\infty$, this leads to a set of coupled mean-field equations
for the local magnetizations $ m_{\mathbf{y}} = (1/L_\tau)\sum_\tau S_{\mathbf{y},\tau}$.
They read
\begin{equation}
 m_{\mathbf{y}} = \tanh \beta~ [ J_{\mathbf{y}} m_{\mathbf{y}} + \sum_{\mathbf{z}} m_{\mathbf{z}} + h]~,
\label{eq:mf}
\end{equation}
where the sum is over all nearest neighbors of site $\mathbf{y}$ and
$h\to 0$ is a very small symmetry-breaking magnetic field which we typically set to $10^{-12}$.
If all $J_{\mathbf{y}}\equiv J_h$, the system
undergoes a (sharp) phase transition at $T_h= J_h+2d$, and if all $J_{\mathbf{y}}\equiv J_l$, it
undergoes the transition at $T_l= J_l+2d$. In the temperature range $T_h > T > T_l$, the phase transition
can therefore be tuned by composition $x$.

The mean-field equations (\ref{eq:mf}) can be solved efficiently in a self-consistency cycle.
Using this approach, we studied systems in one, two, and three space dimensions. The system sizes were
up to L=10000 in 1d, and  up to $L=100$ in 2d and 3d. For each parameter set,
the data were averaged over a large number of disorder realizations.
Details will be given with the individual results below.

\begin{figure}[t]
\includegraphics[width=8.5cm,clip]{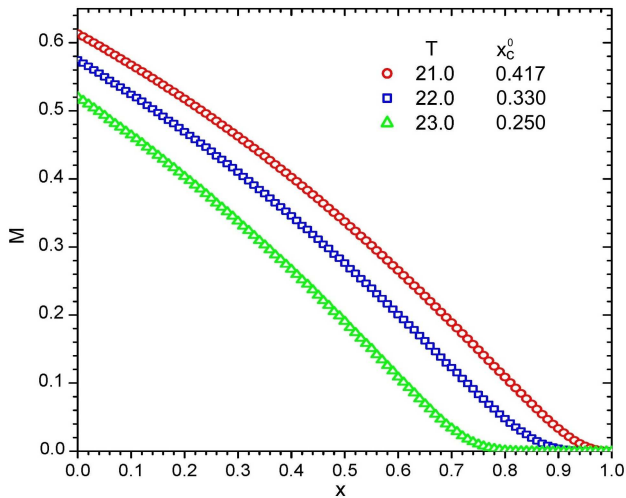}
\caption{(Color online) Magnetization $M$ vs composition $x$ for a $(3+1)$-dimensional
         system having $J_h=20$, $J_l=8$ and several values of the classical temperature
         $T$. The data represent averages over 100 samples of size $L=100$. The values of
         the critical concentration in ``average potential approximation,'' $x_c^0$, are
         shown for comparison.}
\label{fig:overview}
\end{figure}

Figure \ref{fig:overview} shows an overview over the magnetization $M$ as a function of composition $x$
for a $(3+1)$-dimensional system at several values of the classical temperature in the interval
$T_h>T>T_l$.
The figure clearly demonstrates that the magnetic phase extends significantly beyond the
``average potential'' value $x_c^0=(T_h-T)/(T_h-T_l)$. In this sense,
the magnetic phase in our binary alloy benefits from the randomness. In agreement with the smeared phase transition scenario,
the data also show that $M(x)$ develops a pronounced tail towards $x=1$.
(By comparing different system sizes, we can exclude that the tail is due to simple finite-size
rounding.\cite{Vojta03b})
We performed similar simulations for systems in one and two space dimensions,
with analogous results.

\begin{figure}[t]
\includegraphics[width=8.5cm,clip]{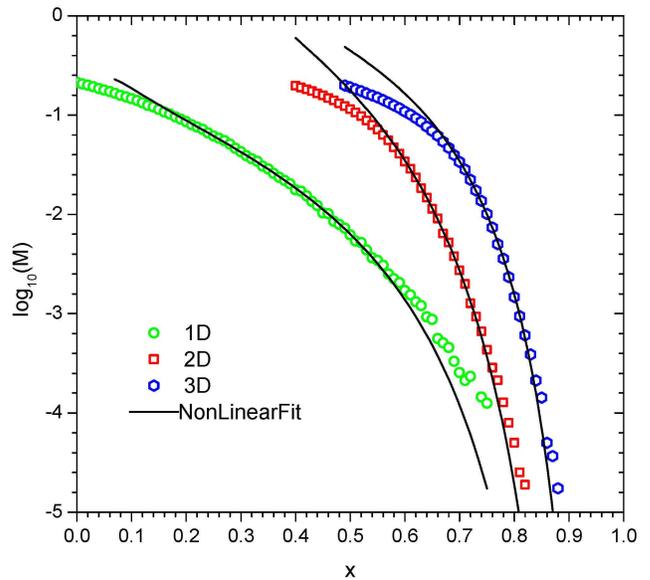}
\caption{(Color online) $\log(M)$ vs $x$ in the tail of the transition for three example systems:
         (i) $(3+1)$-dimensional system with $L=100, J_h=20, J_l=8$, and $T=23$,
         (ii) $(2+1)$-dimensional system with $L=100, J_h=15, J_l=8$, and $T=18$, and
         (iii) $(1+1)$-dimensional system with $L=10000, J_h=11, J_l=8$, and $T=12.8$.
         All data are averages over 100 disorder configurations. The solid lines are fits
         to (\ref{eq:M-exponential}), with the fit intervals restricted to $x\in (0.25,0.55)$
         in (1+1) dimensions, (0.6,0.72) in (2+1) dimensions and (0.7,0.82) for the (3+1)-dimensional
         example.}
\label{fig:exp-fit}
\end{figure}

To verify the theoretical predictions of the optimal fluctuation theory developed
in Sec.\ \ref{sec:OFT}, we now analyze the tail of the smeared phase transition in more
detail. Figure \ref{fig:exp-fit} shows a semi-logarithmic plot of the magnetization $M$
vs.\ the composition $x$ for a $(1+1)$-dimensional system, a $(2+1)$-dimensional system,
and a $(3+1)$-dimensional one.
In all examples, the data follow the theoretical prediction (\ref{eq:M-exponential})
over at least 2 orders of magnitude in $M$ in a transient regime of intermediate
compositions $x$.

\begin{figure}[t]
\includegraphics[width=8.5cm,clip]{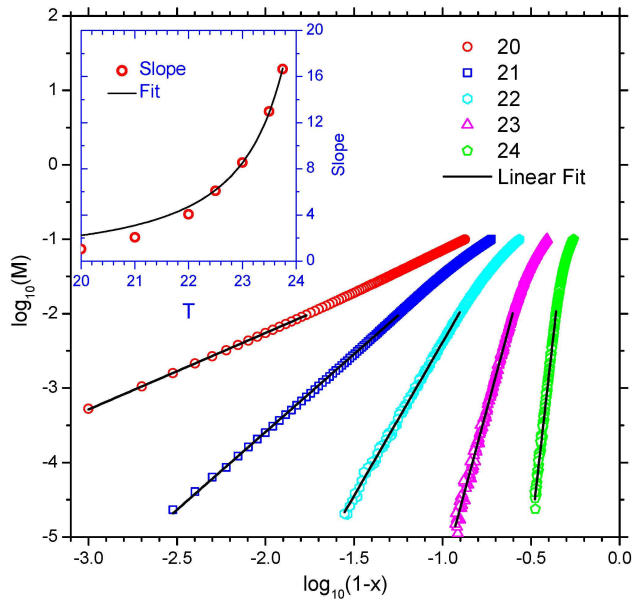}
\caption{(Color online) $\log(M)$ vs $\log(1-x)$ for a $(3+1)$-dimensional
         system with $L=100, J_h=20, J_l=8$ and several temperatures.
         All data are averages over 100 disorder configurations. The solid lines are fits
         to the power-law (\ref{eq:M-power}). The inset shows the exponent
         as a function of temperature, with the solid line being a fit to
         $[x_c^0(T)]^{-3/2}$.}
\label{fig:slopefit}
\end{figure}

We also check the behavior of the magnetization for compositions
very close to $x=1$. Since (\ref{eq:M-power})
predicts a non-universal power law, we plot $\log(M)$ vs.\ $\log(1-x)$ for a $(3+1)$-dimensional
system in Fig.\ \ref{fig:slopefit}.
The figure shows that the magnetization tail indeed decays as a power of $(1-x)$
with $x\to 1$. The exponent increases with increasing temperature in agreement with
the prediction that it measures the minimum size $N_{\rm min} \sim L_{\rm min}^d$ a rare regions needs to have to
undergo the transition independently. The inset of Fig.\ \ref{fig:slopefit} shows
a fit of the exponent to $L_{\rm min}^d \sim [x_c^0(T)]^{-3/2} = [(T_h-T)/(T_h-T_l)]^{-3/2}$. The equation
describes the data reasonably well; the deviations at small exponents can be explained by the
fact that our theory assumes the rare-region size to be a continuous variable which is not fulfilled
for rare regions consisting of just a few lattice sites.

Our computer simulation thus confirm the theoretical predictions in both composition
regions in the tail of the transition. In a transient regime above $x_c^0$, we observe
the exponential dependence (\ref{eq:M-exponential}) while the magnetization for $x\to 1$
follows the non-universal power law (\ref{eq:M-power}).

\section{Conclusions}
\label{sec:conclusions}

In summary, we have investigated phase transitions that are tuned by changing the composition $x$ in
a random binary alloy A$_{1-x}$B$_x$ where pure A is in the ordered phase while pure B is in
the disordered phase. If individual, rare A-rich spatial regions develop true static order, they can
be aligned by an infinitesimal residual interaction. This results in the smearing of the global phase transition,
in agreement with the classification put forward in Ref.\ \onlinecite{VojtaSchmalian05}.

As an example, we have studied the quantum phase transition of an itinerant Ising magnet of the type
A$_{1-x}$B$_x$. At zero temperature, the ordered phase in this binary alloy extends over the entire
composition range $x<1$, illustrating the notion of a smeared quantum phase transition. Upon raising
the temperature, a sharp phase transition is restored, but the transition temperature $T_c(x)$ is
nonzero for all $x<1$ and reaches zero only right at $x=1$ (see Fig.\ \ref{fig:pd}). Using optimal fluctuation theory,
we have derived the functional forms of various thermodynamic observables in the tail of the smeared
transition.
We have also briefly discussed smeared classical phase transitions that can occur in systems with
correlated disorder, and we have performed computer simulations of a toy model that confirm and illustrate
the theory.

Although our results are qualitatively similar to those obtained for smeared phase transitions occurring
at fixed randomness as a function of temperature or an appropriate quantum control parameter,
the functional forms of observables are not identical. The most striking difference can be found in the far tail
of the transition. In the case of composition-tuning, the order parameter vanishes as a
non-universal power of the distance from the end of the tail ($x=1$), reflecting the fact that
the minimum rare region size required for local magnetic order is finite. In contrast, if the transition
occurs at fixed composition as a function of temperature or some quantum control parameter, the order parameter vanishes
exponentially,\cite{Vojta03a,Vojta03b} because the minimum size of an ordered rare region diverges
in the far tail.
These differences illustrate the fact that the behavior of observables
at a smeared phase transition is generally \emph{not} universal in the sense of critical phenomena;
it depends on details of the disorder distribution and how the transition is tuned. Only the question of
whether or not a particular phase transition is smeared is universal, i.e., determined only by symmetries
and dimensionalities.

Let us briefly comment on the relation of our theory to percolation ideas. The optimal fluctuation theory
of Sec.\ \ref{subsec:OFT} applies for compositions $x$ larger than the percolation threshold of the A-atoms.
Because the A-clusters are disconnected in this composition range, percolation of the A atoms does not play
a role in forming the tail of the ordered phase at large $x$. Instead, distant rare regions are coupled via
the fluctuations of the paramagnetic bulk phase and, in metallic magnets, via the RKKY interaction. Percolation does play
a role, though, in the crossover between the inhomogeneous order in the tail of the transition and the bulk order at lower $x$.

We note in passing that the behavior of a diluted system (where B represents a vacancy) with \emph{nearest-neighbor}
interactions is not described by our theory. In this case, the A-clusters are not coupled at all for compositions
$x$ larger than the A percolation threshold. Therefore they cannot align, and long-range order is impossible.
As a result, the super-paramagnetic behavior of the locally ordered clusters extends all the way down to zero
temperature. This was recently discussed in detail on the example of a diluted dissipative quantum Ising model.\cite{HoyosVojta06}

In the present paper, we have assumed that the A and B atoms are distributed independently over the lattice sites,
i.e., we have assumed that there are no correlations between the atom positions. It is interesting to ask
how the results change if this assumption is not fulfilled, for example because like atoms tend to cluster.
As long as the correlations of the atom positions are short-ranged (corresponding to a finite, microscopic
length scale for clustering), our results will not change \emph{qualitatively}. All arguments in the optimal fluctuation theory
still hold using a typical cluster of like atoms instead of a single atom as the basic unit. However, such
clustering will lead to significant \emph{quantitative} changes (i.e., changes in the non-universal
constants in our results), as it greatly increases the probability
of finding large locally ordered rare regions. We thus expect that clustering of like atoms will
enhance the tail and move the phase boundary $T_c(x)$ towards larger $x$.
A quantitative analysis of this effect requires explicit information about the type of correlations
between the atom positions and is thus relegated to future work.

Let us finally turn to experiment. Tails of the ordered phase have been observed at many quantum phase transitions.
However, it is often not clear whether these tails are an intrinsic effect or due to experimental difficulties
such as macroscopic concentration gradients or other macroscopic sample inhomogeneities. Recent highly sensitive
magneto-optical experiments on Sr$_{1-x}$Ca$_x$RuO$_3$ have provided strong evidence for a smeared ferromagnetic
quantum phase transition.\footnote{L. Demko et al., unpublished.} The behavior of the magnetization and critical
temperature in the tail of the smeared transition agree well with the theory developed here. Moreover,
the effects of clustering discussed above may explain the wide variation of the critical composition
between about 0.5 and 1 reported in earlier studies.\cite{CMSCG97,YIKTHK99,KYKMG99}
We expect that our smeared quantum phase transition scenario applies to a broad class of itinerant systems with
quenched disorder.

\section*{Acknowledgements}

We thank I. Kezsmarki for helpful discussions.
This work has been supported in part by the NSF under grant no. DMR-0906566.

\bibliographystyle{apsrev4-1}
\bibliography{../00Bibtex/rareregions}
\end{document}